\pdfoutput=1
\documentclass[aps,pre, twocolumn, groupedaddress]{revtex4-1}

\usepackage{lipsum}
\usepackage{mathtools}
\usepackage{graphicx}
\usepackage{dcolumn}
\usepackage{amsmath}    
\usepackage{amssymb}
\usepackage{bm}
\usepackage{hyperref}
\usepackage{latexsym}
\usepackage{verbatim}
\usepackage[normalem]{ulem}
\usepackage{color}
\usepackage[caption=false]{subfig}
\def\beq{\begin{equation}}
\def\eeq{\end{equation}}

\def\beq{\begin{equation}}                          
\def\eeq{\end{equation}}                          
\def\bea{\begin{eqnarray}}                          
\def\eea{\end{eqnarray}}

\DeclareRobustCommand{\uvec}[1]{{%
  \ifcsname uvec#1\endcsname
     \csname uvec#1\endcsname
   \else
    \bm{\hat{\mathbf{#1}}}%
   \fi
}}
\preprint{}
\begin{document}
\preprint{}
%%%%%%%%%%%%%%%%%%%%%%%%%%%%%%%%%%%%%%%%%%%%%%%%%%%
%                               TITLE & ABSTRACT
%%%%%%%%%%%%%%%%%%%%%%%%%%%%%%%%%%%%%%%%%%%%%%%%%%%
%Title of paper
\title{Ordering kinetics  in active polar fluid}
\author{Shambhavi Dikshit$^{1}$}
\email{shambhavidikshit.rs.phy18@itbhu.ac.in}
\author{Shradha Mishra$^{1}$}
\email{smishra.phy@iitbhu.ac.in}
\affiliation{$^{1}$Indian Institute of Technology (BHU), Varanasi, India 221005}
\date{\today}
\begin{abstract}
We model the active polar fluid as a collection of orientable objects
supplied with active stresses and momentum damping
coming from the viscosity of bulk fluid medium. 
The growth kinetics of local orientation  field is studied. The effect of active fluid is contractile or extensile depending upon the sign of the active stress. We explore the growth kinetics for different activities.
We observe that for both
extensile and contractile cases the growth is altered by a
prefactor when compared to the equilibrium {\em Model A}.
We find that the extensile fluid enhances the domain
growth whereas the contractile fluid supresses it. The
asymptotic growth becomes pure algebraic for large
magnitudes of activity. 
We also find that the domain morphology remains unchanged due to activity and system shows the good dynamic scaling for all activities. Our study provides the understanding of ordering kinetics in active polar gel.
\end{abstract}
\maketitle
 The systems in which  the energy consumption occurs on individual constituent level and it leads to collective dynamics are active systems \cite{Ramaswamy2010, Marchetti2013}. The existence of active systems is found from small microscopic length scale, i.e. interacellular level like cytoskeletal actin filaments \cite{Rappel1998}, bacterial colonies \cite{Berg2004} etc. to large macroscopic scale i.e. upto few meters like animal herds \cite{Couzin2005}, birds flocks \cite{Vicsek1995, Cavagna2008} etc. Active systems are defined as wet when coupled to a momentum conserving solvent, in which solvent mediated hydrodynamic interaction becomes important \cite{Ramaswamy2010, Marchetti2013}. Bacterial swarms in a fluid, cytoskeleton filaments, colloidal or nanoscale particles propelled through a fluid are examples of wet systems \cite{Paxton2004}, \cite{Bechinger2016}.  When no such fluid is present, then system is called dry. Dry systems include bacteria gliding on a surface \cite{Wolgemuth2008}, animal herds or vibrated granular particles and so on \cite{Toner1998, Ramaswamy2003}.           \\
Starting with the seminal work of Vicsek \cite{Vicsek1995}, most of the previous works on active system have focused on the steady state properties  \cite{Ramaswamy2005, Chate2008, Mishra2012, Marchetti2013}.
 The study of ordering kinetics in active systems is complex by the fact that
the system relaxes to a nonequilibrium steady state (NESS). There have been very few studies \cite{Das2018, Mishra2014, Pattanayak2021, Wittkowski2014} of the
coarsening kinetics from a homogeneous initial state to
the asymptotic NESS, though understanding it  is of great experimental interest.  Previous studies of coarsening or domain growth have primarily focused upon systems approaching to an equilibrium state \cite{Bray1994, Puri2008, LIFSHITZ1961, Hohenberg1977}. Based on the symmetry and conservation laws the domain growth is classified mainly of two types. The domain growth in systems  with conserved order parameter is named as {\em Model B} and with nonconserved order parameter is called as {\em Model A},  follows an algebraic growth law with growth exponent $z = 3$ \cite{LIFSHITZ1961} and $2$ \cite{Hohenberg1977} respectively. For the systems with scalar order parameter and nonconserved growth kinetics, the interfacial velocity of the growing domain is proportional to the local curvature of the interface; that leads to the size of the domain $L(t) \propto t^{1/2}$; Allen and Cahn growth law \cite{Allen1979}. Whereas for the systems with conserved kinetics the interface have to pay a cost due to local conservation of order parameter. That leads to the size of the domain grows with time such that $L(t) \propto t^{1/3}$; Lifshitz-Slyozov-Wagner (LSW) theory \cite{LIFSHITZ1961,Wagner1961}. \\
For the systems having symmetries of two-dimensional {\em XY-}model, with nonconserved growth kinetics and order parameter with more than one components or vector order parameter, the asymptotic growth law is still $z=2$. The topological defects  are vortices and antivortices and the domain growth is driven by the annihilation of these defects. The detailed calculation \cite{Blundell1994,Goldenfeld1990} shows that there is logarithmic correction to the pure algebraic growth $L(t) \propto (t/\ln(t))^{1/2}$.
Equilibrium liquid crystals,  ferromagnetic materials with continuous symmetry, spin glasses, two-dimensional superconductors, etc. are some of the examples of  systems with nonconserved  vector order parameter. \\
The domain growth in systems approaching towards a thermal equilibrium state is very well studied in dry \cite{LIFSHITZ1961, Hohenberg1977, Blundell1994, Goldenfeld1990} as well as wet systems with hydrodynamic effect \cite{Tiribocchi2015, Navarro, Alarcon, Llopis}. 
 The understanding of the ordering kinetics in terms of the number of topological defects is explored in many of the studies \cite{Julicher2007,Simha2002,Elgeti2011,Thampi2014,giomi2015}. Recently some studies are performed on the understanding of ordering of domain growth   in {\em dry} active systems \cite{Pattanayak2021,Pattanayak21,Das2018, Wittkowski2014}. But the ordering kinetics in active systems {\em with fluid}  is rarely explored. Some recent studies show the effect of hydrodynamics on the  ordering kinetics of apolar order parameter field \cite{Kumar2022} and  the effect of fluid on the steady state properties of active polar fluid \cite{Voituriez, Marenduzzo, Kruse2004}. This motivates us to study the ordering kinetics 
of active polar systems with fluid or active polar gel. The examples of active polar gels are bacteria suspensions, active emulsions and active gels. \cite{Saha2022, Prost2015}.\\
The model contains a
collection of orientable objects
supplied with active stresses and momentum damping coming from the viscosity of bulk fluid medium.
The ordering kinetics of the orientational field is studied after a quench from the random  disordered state. With time the system orders and the size of ordered domain grows with time. We characterise the domain growth and scaling. The direction of  spontaneous flow makes the system to respond like  extensile and contractile in nature. For the extensile case, particles act like pushers (pulling fluid inward equatorially    and
emitting it axially) and for contractile case they are more likely pullers \cite{Cates2011} (vice versa).  We observe that for both extensile and contractile cases  the  growth  is altered by a prefactor when compared to the equilibrium {\em Model A}. 
                       We find that the extensile fluid enhances the domain growth whereas the contractile fluid supresses it. The asymptotic growth becomes pure algebraic for large magnitudes of activity. For all activities,  system shows good dynamic scaling and domain morphology remains unaffected with respect to the activity.\\

{\em Model A}:-  
The time evolution of  system with nonconserved local order parameter for a collection of orientable objects is describe by the time-dependent Ginzburg Landau equation  \cite{Lubenskey,Hohenberg1977}

\begin{equation}
	  \frac{\partial  {\textit{P}} _{\alpha}({\bf r},t)}{\partial t}= -\Gamma_{0}\frac{\delta F_{0}}{\delta  {\textit{P}}  _{\alpha}({ \bf r},t)}+\theta_{\alpha}(\textbf{r},t)
   \label{eq:1}	  
  \end{equation}
                where $\Gamma_{0}$ is the mobility. The Ginzburg Landau free energy  $F_{0}$  is 
\begin{equation}
       F_{0}= \int d^{d}r\{ {\frac{1}{2}  a{ \textbf {\textit{P}}}({\bf r},t) ^{2}+\frac{\lambda}              {2}|\nabla{\textbf {\textit{P}}}({\bf r},t) ^{2}|+\frac{b}{4}{\textbf {\textit{P}}}({\bf r},t) ^{4}\}}
       \label{eq:2}
   \end{equation}  

here ${\textbf {\textit{P}}}({\bf r}, t)$ is a vector field  with components ${\textit{P}}_{\alpha}(r,t)$, and $\alpha = 1$ and $2$ in two-dimensions.
  The vector field ${\textbf  {\textit{P}}}({\bf r}, t)$  is the local orientation field and is defined by the average orientation of the particles in a small coarse-grained region. The size of the region is such that it  consists of sufficient number of particles to perform the statistical averaging.
  $\theta_{\alpha}({\bf r} ,t)$ is Gaussian random white noise with properties
 $<\theta_{\alpha}({\bf r} ,t)>=0$ and  $<\theta_{\alpha}({\textbf r},t)\theta_{\alpha}({\textbf r} ^{'},t^{'})>=2\Delta_{0}\delta(t-t^{'})\delta_{\alpha\alpha^{'}}$. $a$, $b$ and $\lambda$ are constants. ($a < 0$ ensures the broken symmetry state, $b > 0$ and $\lambda>0$ for stability and the strength of noise $\Delta_0=0$). 
 After substituting the form of $F_0$ from Eq. \ref{eq:2} in Eq. \ref{eq:1}  and performing the functional derivative of Ginzburg Landau free energy $F_0$, we get the time-dependent Ginzburg Landau (TDGL)  \cite{Hohenberg1977} equation for nonconserved vector field
 
\begin{equation}
%\begin{split}
    \frac{\partial {\textbf {\textit{P}}} ({ \bf r},t)}{\partial t}  =a \Gamma_0{ \textbf {\textit{P}}} ({\bf r},t)\\
    -b \Gamma_0 |{\textbf {\textit{P}}} ({ \bf r},t)|^2 {\textbf {\textit{P}}} ({\bf r},t)+\lambda\nabla^2 {\textbf {\textit{P}}}({\bf r},t)
	  \label{eq:eqp}
%\end{split}	  
\end{equation}    
 The model describe by Eq. \ref{eq:eqp}  is called as {\em Model A} according to the Halprin and Hohenberg \cite{Hohenberg1977}. The noise term present in Eq. \ref{eq:1} is {\em turned off} and we consider the deterministic part of the TDGL equation as discussed in \cite{Puri2008}. The Gaussian noise in Eq. \ref{eq:1} is purely thermal in nature and  ensures that the system reaches the global minima at late times. But most of the kinetic theories are developed for the deterministic TDGL equation and thermal noise is irrelevant for growth kinetics \cite{Puri2008}. The equation  \ref{eq:eqp}  very well explains the ordering kinetics in magnets with vector order parameter \cite{Kim1998} and liquid crystals \cite{Orihara1993}. Now we further introduce the effect of hydrodynamic interaction on the ordering kinetics of nonconserved field.\\

\begin{figure*}[ht]
\centering
     \includegraphics[scale=0.25]{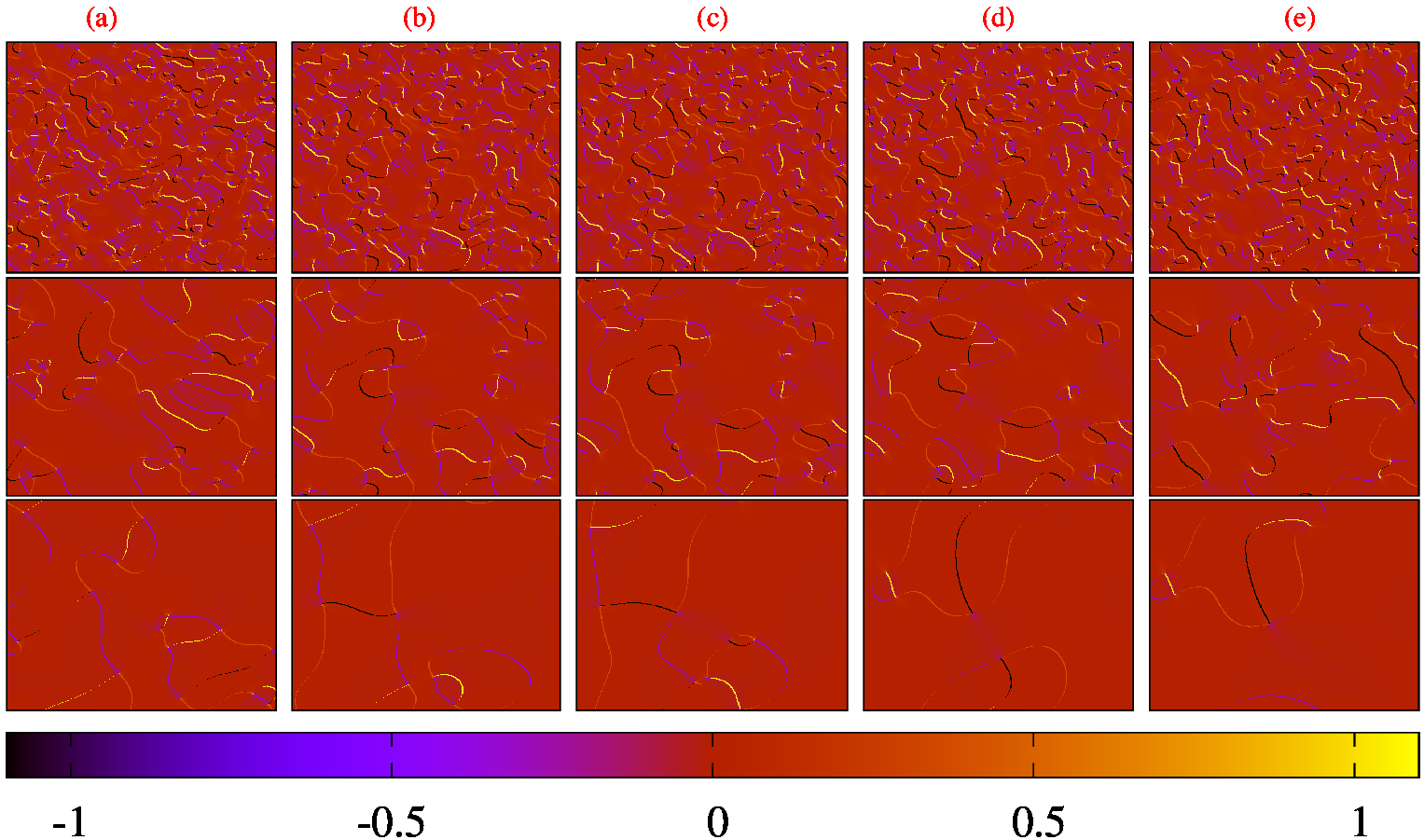}
      \includegraphics[scale=0.255]{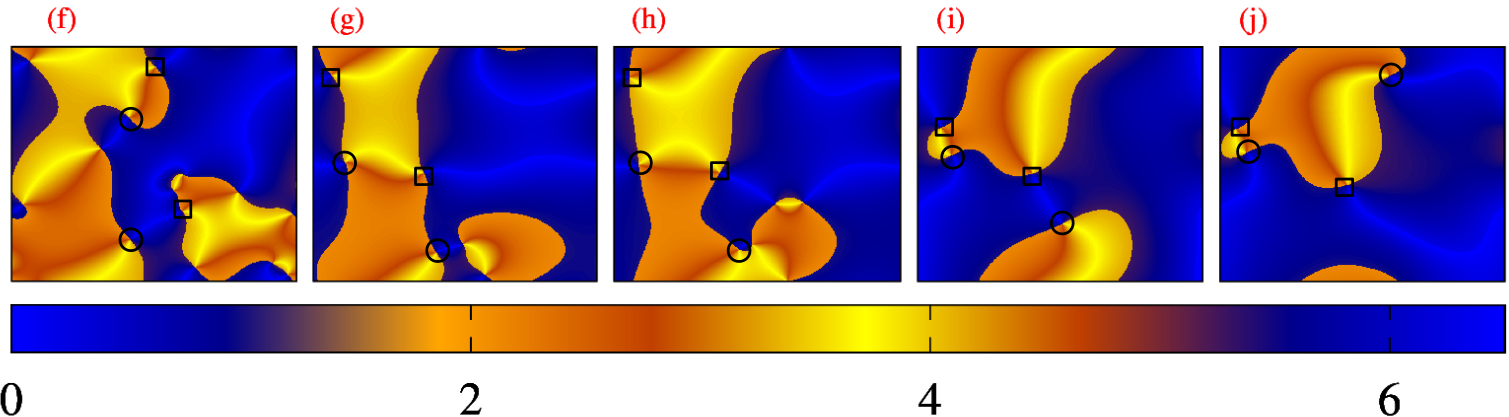}
       \includegraphics[scale=0.245]{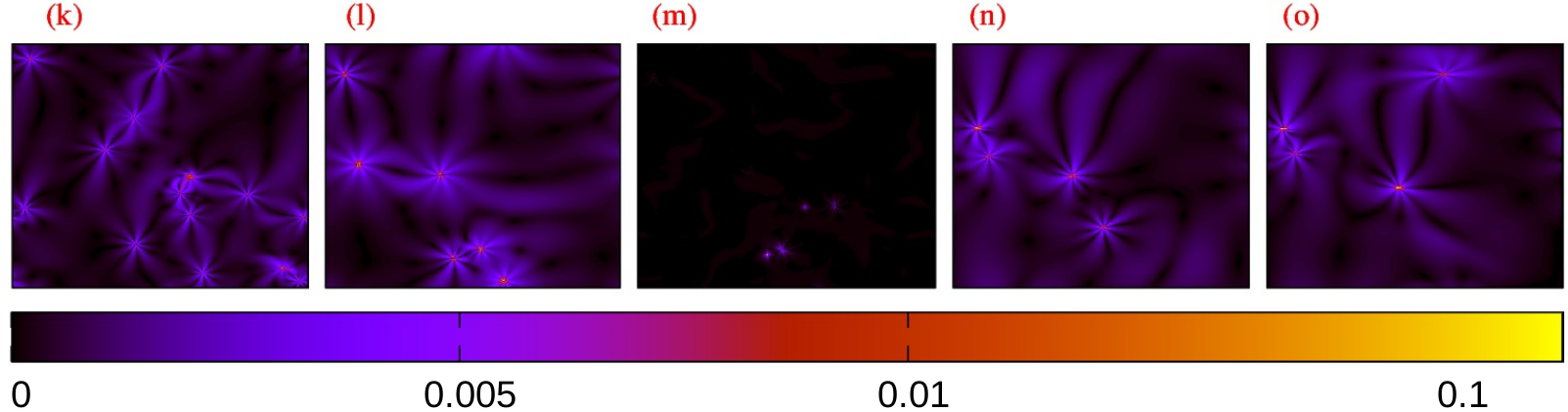}
	\caption{(color online)(a)-(e) are plot of left hand size of equation \ref{eq:14} for different activities $-3,-1,0,1$ and $3$ respectively. Upper to bottom panels are for different times $80, 800$ and $8000$.  (f)-(j) and (k-o) are corresponding angle $\theta({\bf r}, t)$ and magnitude of fluid velocity $|{\bf v}({\bf r}, t)|$ respectively for the same set of activities as in (a-e) and  at late time $8000$. In (f)-(j) the circles and squares represent the location of some of the vortices and antivortices with winding number $k= +1$ and $-1$ respectively.} 
	\label{fig:fig1}
\end{figure*}                   

\begin{figure}[ht]
\centering
\includegraphics[scale=0.20]{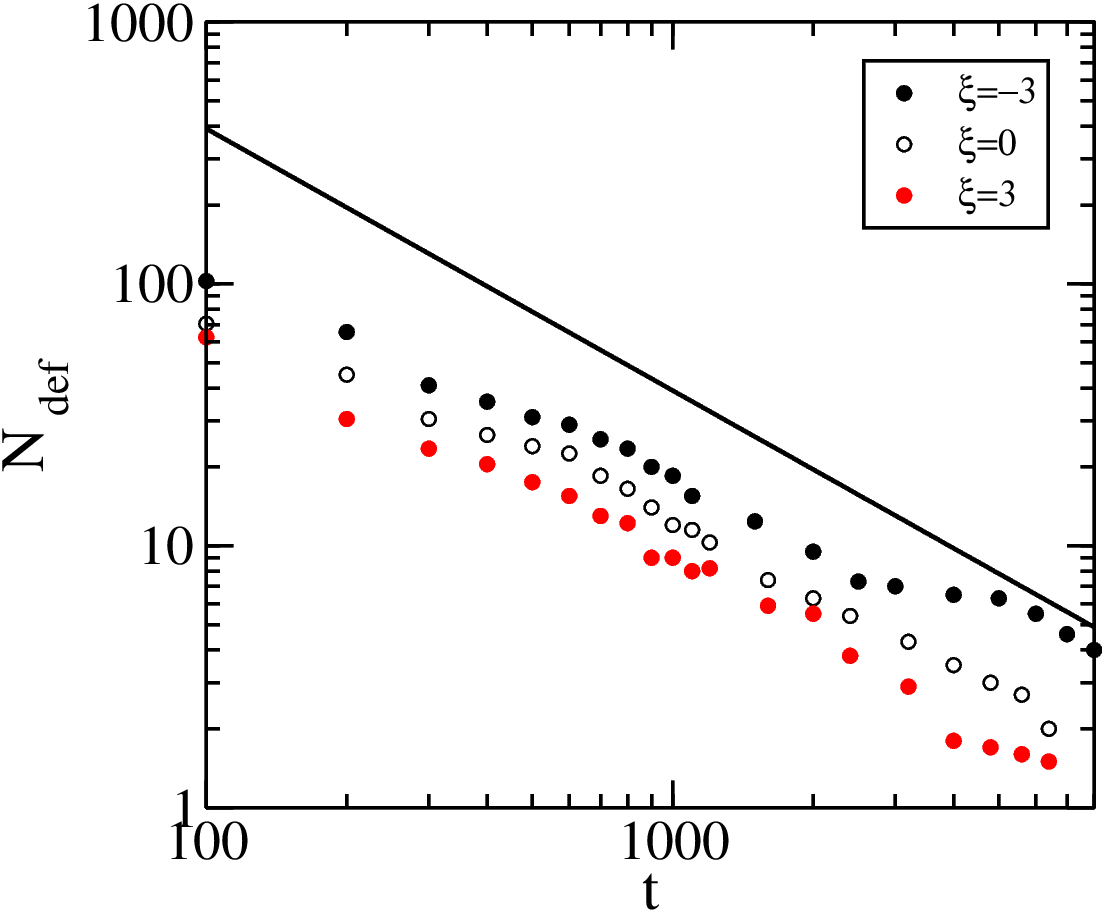}
 \caption{(color online) Number of defects vs. time is plotted for three different activities  $-3,0,3$. Black solid line of slope $-1$ is drawn.}
  \label{fig:fig2}
\end{figure} 

 {\em Active Polar Fluid}:-
   Now we discuss the hydrodnyamics of  active polar fluid or polar active poalr gel. We focus here the active gel defined with collection of orientable objects supplied with active stresses and momentum damping coming from the viscosity of bulk fluid medium, compare to the friction due to substrate or medium \cite{Simha2002, Julicher2007, Kruse2004}. The equations are first proposed by \cite{Simha2002} for self-propelling objects but later developed by \cite{Kruse2004} and \cite{Julicher2007} for the system of cytoskeleton in living cells and polar actin filament, which become active only in the presence of molecular motors that consumes ATP.  A collection of artificial Janus rods which gain motility due to electrophoresis is a good example of active polar gel and can be easily designed in the laboratory \cite{Paxton2004}. The presence of fluid introduces the hydrodynamic effect. If the hydrodynamic interaction is turned off then the model is purely passive and same as {\em Model A}. The model incorporate the coupling between local order parameter and fluid. \\
   We model the system by the coupled  dynamics of the orientation order parameter ${\textbf {\textit{P}}}({\bf r}, t)$ with a solvent local velocity ${\textbf {\textit{v}}}({\bf r}, t)$ with additional active stresses. 
    The system is modeled by the coarse-grained 
coupled hydrodynamic  non-linear partial differential equations of motion for the two fields. The fluid is introduced through the standard Navier-Stokes equation of motion for fluid with additional coupling to the polarisation, ${\textbf {\textit{P}}}({\bf r}, t)$ of 
particle through active and passive stresses (deviatoric stress) as introduced in \cite{Kruse2004}. \\
In the presence of fluid, in addition to the term present in Eq. \ref{eq:eqp} coupling to the fluid velocity, hence the Eq. \ref{eq:eqp} will have 
 convective nonlinearity of type
 $({\textbf {\textit{v}}}.\nabla) {\textbf {\textit{P}}}$.
 Hence the modified equation for the  ${\textbf {\textit{P}}}({\bf r}, t)$ will become
\begin{equation}
\begin{split}
    \frac{\partial {\textbf {\textit{P}}} ({ \bf r},t)}{\partial t} +{\textbf {\textit{v}}} .\nabla {\textbf {\textit {P}}} ({ \bf r} ,t)+\omega_{\alpha\beta}P_{\beta}+v_{1}v_{\alpha\beta}P_{\beta} =a \Gamma_0{ \textbf {\textit{P}}} ({\bf r},t)\\
    -b \Gamma_0 |{\textbf {\textit{P}}} ({ \bf r},t)|^2 {\textbf {\textit{P}}} ({\bf r},t)+\lambda\nabla^2 {\textbf {\textit{P}}}({\bf r},t)
	  \label{eq:3}
\end{split}	  
\end{equation}
 with the comoving and corotational derivative of the polarisation,  ${\textbf {\textit{P}}}({\bf r}, t)$, 
  where $\omega_{\alpha\beta}=\frac{1}{2}(\partial_{\alpha}v_{\beta}-\partial_{\beta}v_{\alpha})$ and $v_{\alpha\beta}=\frac{1}{2}(\partial_{\alpha}v_{\beta}+\partial_{\beta}v_{\alpha})$ are the vorticity and strain-rate tensor respectively.
   The coupled velocity field is due to momentum conserving solvent which satisfies the condition of incompressibility, {\em i.e.} 
%\begin{equation}
         $\nabla. \textbf {\textit{v}} =0$
          % \label{eq:4}
%\end{equation}
     
           The equation for  fluid velocity, ${\textbf {\textit{v}}} ({\bf r} ,t)$,  satisfying conservation of mass (condition of incompressibility) and conservation of momentum is given as 
\begin{equation}
	  \frac{\partial {\textbf {\textit{v}}}}{\partial t}+{\textbf {\textit{v}}.\nabla} {\textbf {\textit{v}}}=\eta\nabla^2{\textbf {\textit{v}}} -\nabla p+\nabla.\sigma_{\alpha\beta}^{total}
	  \label{eq:5}
\end{equation} 

 The Eq. \ref{eq:5} is the Navier-Stokes equation with an additional force term due to stresses present in the fluid. This term  includes fluid static part and fluid dynamic part that involves activity and flow coupling coefficient.
In this case total stress tensor becomes 
$ \sigma_{\alpha\beta}^{total}= \sigma_{\alpha\beta}+ \sigma_{\alpha\beta}^a$
where $\sigma_{\alpha\beta}$ is fluid passive
part and $\sigma_{\alpha\beta}^a$ gives fluid active part.
In polar liquid, mechanical stress tensor can be decomposed into symmetric and antisymmetric part, where the symmetric part of stress tensor is the actual thermodynamics flux and conjugate force is the antisymmetric part of the velocity gradient $v_{\alpha\beta}$. 
 Hence the constitutive equation for  stress tensor gives \cite{Kruse2005}
\begin{equation} 
  \sigma_{\alpha\beta}^{total}=2\eta_{1}v_{\alpha\beta}+\frac{1}{2} v_{1}(P_{\alpha}h_{\beta}+P_{\beta}h_{\alpha}-\frac{d}{2}(P_{\gamma}h_{\gamma}\delta_{\alpha\beta}))+\xi q_{\alpha\beta}
  \label{eq:7}
\end{equation}
here $d=2$ for the two-dimensions. 
$\xi$ is the transport coefficient related to the activity of the system. The term $\xi q_{\alpha\beta}$ is self-propelled stress, first incorporated by \cite{Ramaswamy2002} into generalised hydrodynamics of orientable fluid. 
The sign of the activity coefficient, $\xi$ tend to change the nature of the system. A negative  of $\xi$ corresponds to a contractile stress as in the polar active filament \cite{Kruse2005, Kruse2004}. A positive $\xi$ shows an  extensile stress as observed in certain Bacterial suspensions \cite{Marchetti2013}. 
$v_{1}$ is the flow coupling coefficient and  $q_{\alpha\beta}=P_{\alpha}P_{\beta}-\frac{1}{d}\delta_{\alpha\beta}$.
{If the coefficient $\xi=0$ turns off the hydrodynamic coupling is purely passive,  which represents the {\em Model A} with fluid or  we called it {\em Passive Model A}}.

\begin{figure*} [ht]
\centering
\includegraphics[scale=0.29]{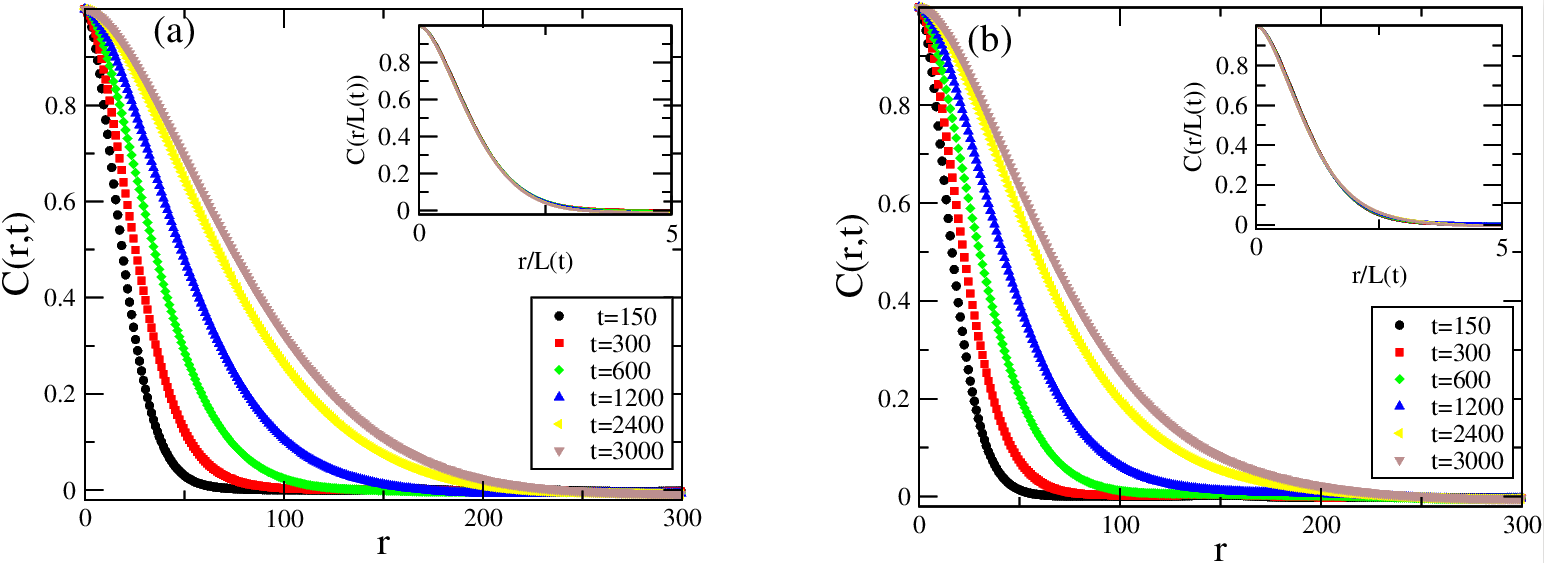}
	\caption{(color online) Two-point correlation function $C(r, t)$ {\em vs.} distance $r$ at different times for two different $\xi$ = $3$ and $-3$ for system size $L=1024$ in (a) and (b) respectively. The inset figures show the scaled two-point correlation $C(r/L(t))$ {\em vs.} scaled distance $r/L(t)$.}
\label{fig:fig3}
\end{figure*} 

The pressure term on the right hand side of Eq. \ref{eq:5} can be eliminated by taking the $\it curl$ $(\times)$  on both sides of Eq. \ref{eq:5} we find the equation for the vorticity of fluid $\omega =  \nabla\times {\bf v}$
 \begin{equation}
\frac{\partial \omega}{\partial t}+({\textbf  {\textit{v}}}.\nabla)\omega=\eta\nabla^2\omega+\nabla\times(\nabla.\sigma_{\alpha\beta}^{total}) 
   \label{eq:6}
\end{equation}
by integrating Eq. $\ref{eq:6}$, we get $\omega$ and then we  solve Poisson's equation, 
\begin{equation}
\nabla^2\psi=-\omega
\label{eqomega}
\end{equation}
 where a scalar field $\psi$ is defined such that 
\begin{equation}
{\textbf {\textit{v}}} = (\partial_y \psi, -\partial_x \psi)
\label{eqv}
\end{equation}Then the updated flow field ${\textbf {\textit{v}}}(r,t)$ enters in Eq. $\ref{eq:3}$. 
 We study the ordering kinetics of active polar fluid when quenched from the random disordered state to ordered state. Later everywhere the time and length scales are rescaled by
 $({a \Gamma_0})^{-1}$  and $\sqrt{\frac{\lambda}{a \Gamma_0}}$ respectively to make the  equations and parameters dimensionless.
     
We numerically integrate the Eqs. \ref{eq:3}, \ref{eq:6} and \ref{eqomega} using Euler's scheme with small
steps $\Delta x=1.0$ and $\Delta t =0.1$.  In our numerical implementation, the first-and
second-order derivatives for an arbitrary function $f (r, t)$
are discretized as
\begin{equation}
\frac{\partial f}{\partial t} = \frac{f(t+\Delta t)-f(t)}{\Delta t}    
\end{equation}

\begin{equation}
\frac{\partial f}{\partial x} = \frac{f(x+\Delta x)-f(x-\Delta x)}{2\Delta x}
\end{equation}

\begin{equation}
\frac{\partial^2 f}{\partial x^2} = \frac{f(x+\Delta x)-2f(x)+f(x-\Delta x)}{\Delta x^2}
\end{equation}

We fix the values of coefficients $a$,  $b$, $\lambda$,  $\eta$, $\Gamma_0$ and $v_{1}$
to $1$ and tune the activity $\xi$. The activity $\xi$ is tuned from $[-4, 8]$ to see the effect of both contractile and extensile stresses generated due to fluid present.
Systems is started from the random homogeneous state of orientation of polarisation and random initial scalar field in small range $\psi \in [1.0 - 1.1]$ and then  initial fluid velocity is generated by using Eq. \ref{eqv}. After that  we calculate the vorticity $\omega$ by taking the {\em curl} ($\nabla \times$) of velocity ${\bf v}$. Finally using Eqs. \ref{eq:3} and \ref{eq:6} we updated the local polarisation ${\bf P}({\bf r}, t)$ and vorticity $\omega$ respectively. The further scalar field $\psi$ and velocity ${\bf v}$ is updated using Poisson's Eq. \ref{eqomega} and Eq. \ref{eqv} respectively. This whole process of updates of local polarisation ${\bf P}$ and local fluid velocity ${\bf v}$ is counted as one simulation step. We let the system evolve for total time steps of $t=8 \times 10^4$ (the total real time $t=8 \times 10^3$) and system size $L \times L = 512 \times 512$ and $1024 \times 1024$ with periodic boundary conditions in both the directions.   
The data is averaged over the $25$ independent realisations for good statistics. We checked the numerical stability of the system for the present set of parameters.\\

\begin{figure*}[ht]
\centering
     \includegraphics[scale=0.165]{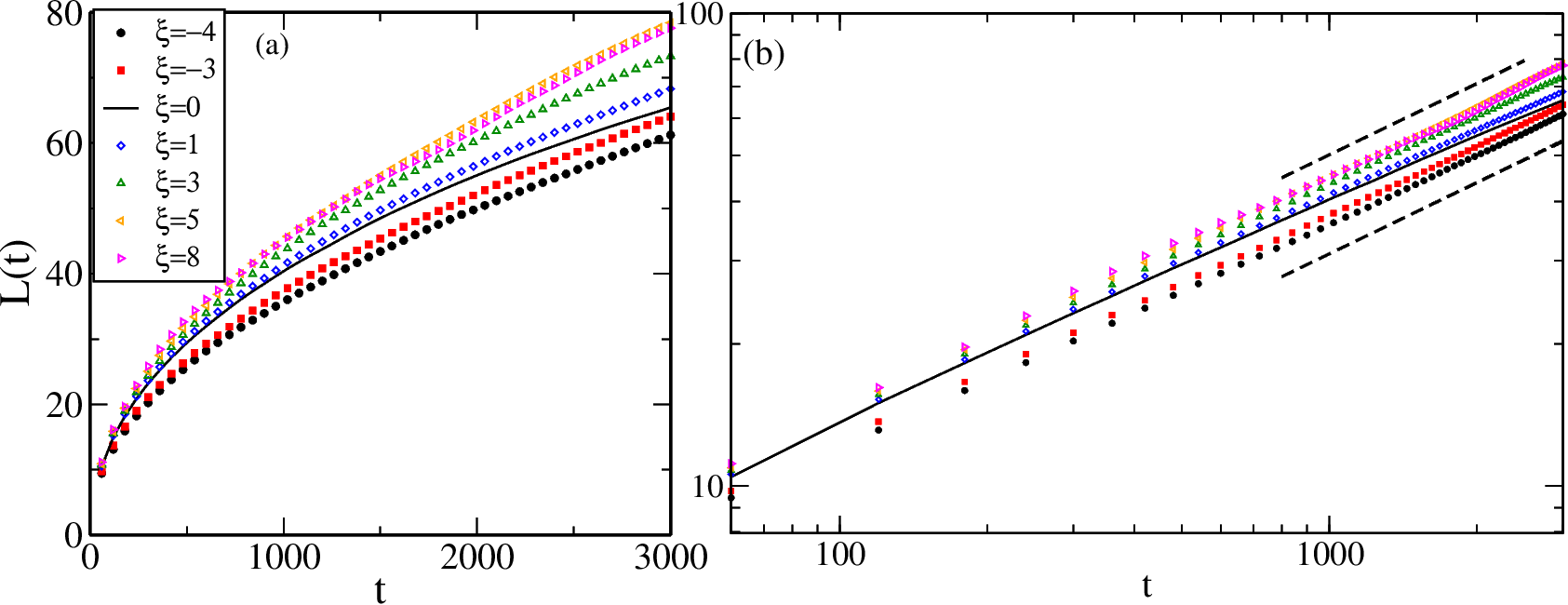}
  \includegraphics[scale=0.12]{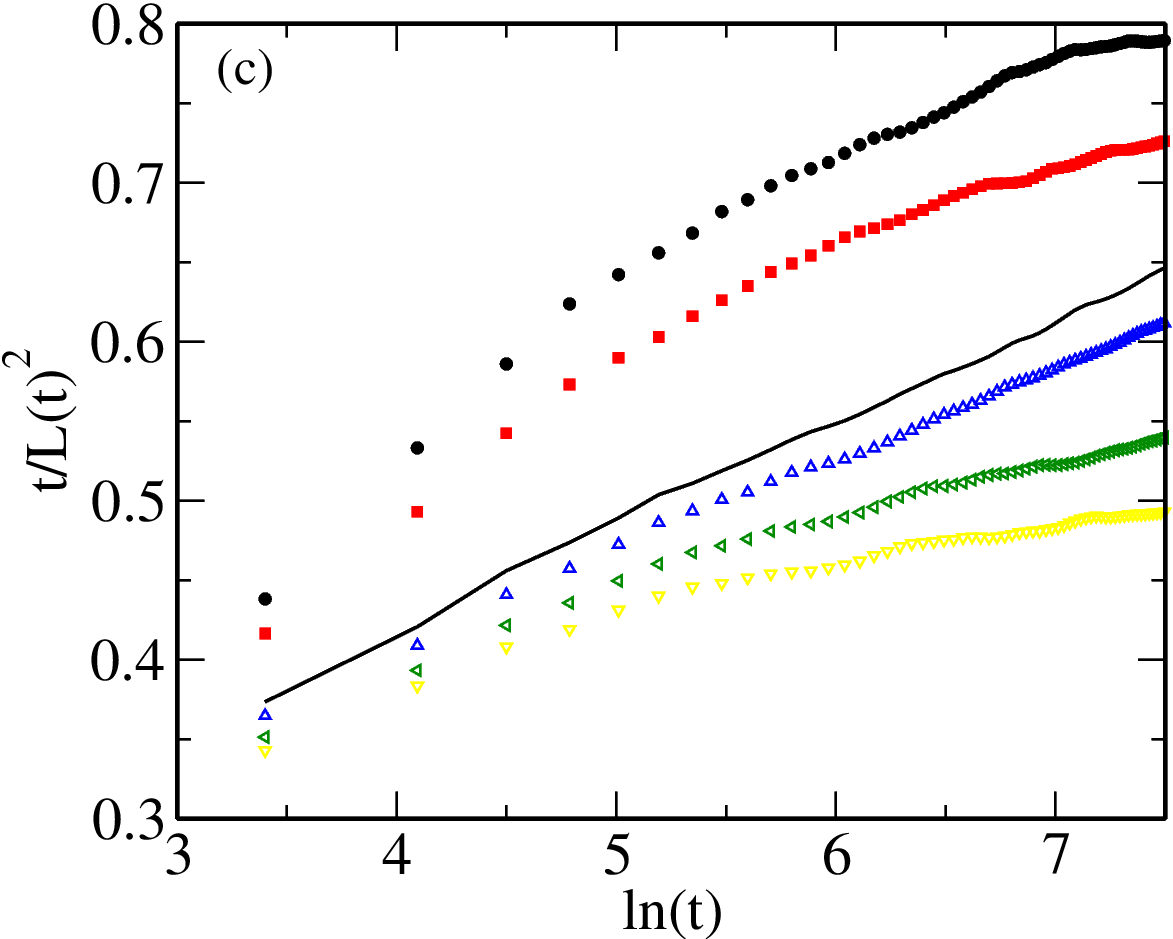} 
	\caption{(color online)(a) Shows the plot of  characteristic lengths $L(t)$ {\em vs.} time $t$ on linear scale for positive $\xi$ =  $0$, $1$, $3$, $4$ and $5$ and for negative $\xi$ =$-3$ and $-4$. (b) shows the same plot on $\log-\log$ scale. The two dahsed lines are lines of slope $1/2$. (c) The plot of $\frac{t}{(L(t)^2)}$ {\em vs.} $\ln(t)$ for the  $\xi$'s as in (a) and (b).}
\label{fig:fig4}
\end{figure*} 

{\em Results}:-  
We first let the system evolve to the ordered state after a quench from the disordered initial state. After the quench, the point-like defects or disclinations are observed. These defects are spatially inhomogeneous configuration of the director field or orientational order field in our system. The strength of a disclination depends on the rotation of orientation field around the defect core in one loop. For two dimensional system, the rotation of director field can be expressed in terms of a single scalar field, $\theta$, representing the angle formed by the director $n=(cos \theta,sin \theta)$ with the horizontal axis of the Cartesian frame. This gives
\begin{equation}
\frac{1}{2 \pi} \oint d \theta= k
\label{eq:14}
\end{equation}    
where the integral is calculated along an arbitrary contour and $k$ is the winding number. If the contour encloses a defect  then the  winding number $k$ of vortex/antivortex disclination is $k= +1$ and $-1$ respectively. For other places it should be almost zero. We calculate  the value of {\em k} for the set of values of activities and variation of $k$ on the two-dimensional plane is shown in fig \ref{fig:fig1}(a-e) for $\xi=-3,-1,0,1$ and $3$ respectively  and for different times $t=80$, $800$ and $8000$, starting from upper panel to lower panel. We observe that as the system evolves, the number of defects decreases and larger the value of $\xi$ the more homogeneous configuration of the orientation field ${\bf P}({\bf r}, t)$ is observed. The number of defects decreases by increasing the activity $\xi$. This we confirm by the angle plot $\theta({\bf r}, t) = \tan^{-1}[\frac{P_2({\bf r}, t)}{P_1({\bf r}, t)}]$, shown in the fig \ref{fig:fig1}(f-j), for the same set of $\xi$ as in (a-e) at late time, $t=8000$. The meeting points of dark and bright colors are the location of defects. The circles and squares in Fig. \ref{fig:fig1} (f-j)  show the locations of some of the vortices  and antivortices  with winding number $k=+1$ and $-1$ respectively. Further in Fig. \ref{fig:fig1}(k-o) we plot the magnitude of fluid velocity $v({\bf r}, t) = |{\bf v}({\bf r}, t)|$ for the same set of activities. The structure of fluid is very different for active and passive cases. Very clearly fluid velocity develops  eight-fold symmetric long ranged pattern around the defect cores for active fluid Fig.\ref{fig:fig1}(k, l, n and o).  Such pattern is absent  and magnitude of fluid velocity is zero for {\em passive Model A} Fig. \ref{fig:fig1}(m) . \\
Next, we quantify  the number of defects, $N_{def}$ with time for three different cases, $\xi=-3, 0$ and  $3$. The $N_{def}$ is calculated  by taking the average number of  vortices and antivortices or (counting the number of points where the winding number $k = \pm 1$ and then average is performed over $10$ independent realisations). From fig \ref{fig:fig2}, we see that $N_{def}$, decreases with time as power law $\propto t^{-1}$. Higher the activity $\xi$, lesser the number of defects are observed. Solid line  of  slope $-1$ is drawn to show the power law decay of $N_{def}$ with time. We further explore the ordering kinetics of the active polar gel for the different activities in following sections.

 {\it Dynamic two-point correlation function}:- The nature and evolution of the structure in the orientation field is characterised by calculating the correlations in orientation order parameter field ${\bf P}({\bf r}, t)$, defined as 
$C(r,t)=<\delta {\textbf {\textit{P}}}({\bf r}_{0}+{\bf r},t) \cdot \delta {\textbf {\textit{P}}}({\bf r}_{0},t)>$ , where $\delta {\textbf {\textit{P}}}$ is the fluctuation from mean  and  $<...>$ denotes  averaging over directions, reference positions $r_{0}$ and $25$ independent realisations. As the system coarsens with time,  correlation function increases as shown in Fig. \ref{fig:fig3} for two different $\xi$'s, $3$ and $-3$ (a) and (b) respectively. 
Further  
we define the characteristic length $L(t)$  as the value of $r$ at which the correlation function $C(r, t)$  decreases to $0.1$ of its value at $r=0$. 
                In the insets of Fig. \ref{fig:fig3}(a-b),  we plot the scaled two-point 
                correlations $C(r/L(t))$ vs. scaled distance $r/L(t)$. We find that all the curves for different times, collapse to a single curve for both contractile $\xi = -3$ and extensile $\xi = +3$ systems. Hence for both the cases system shows the good dynamic scaling.\\
                
\begin{figure*}[ht]
\centering

  \includegraphics[scale=0.24]{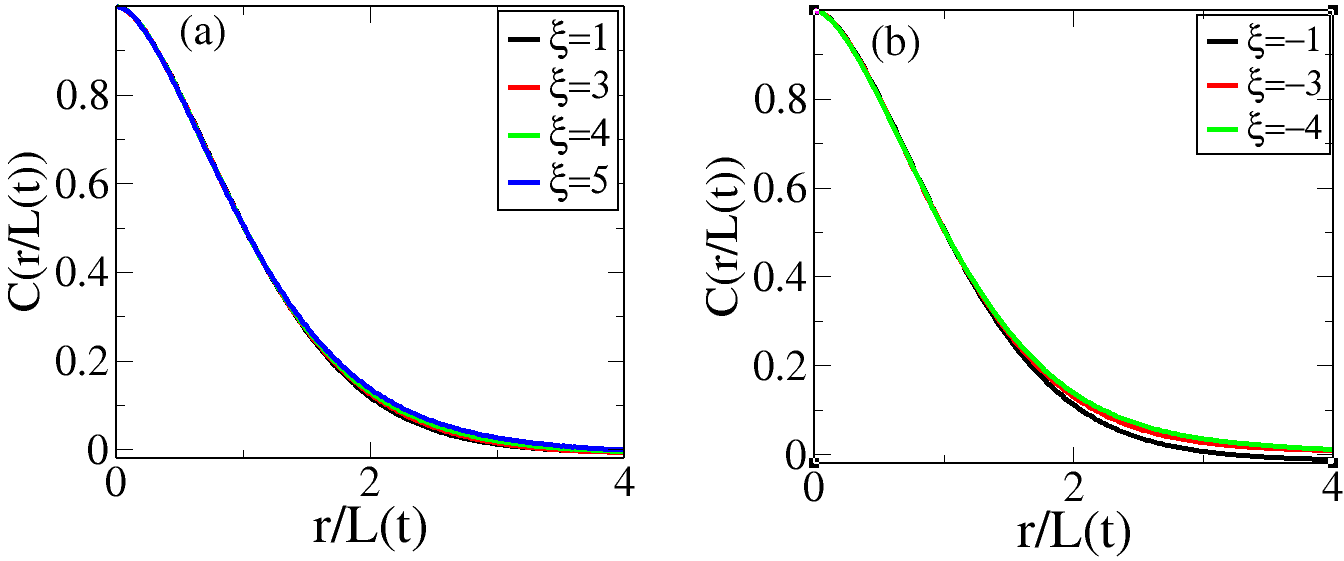}
   \includegraphics[scale=0.15]{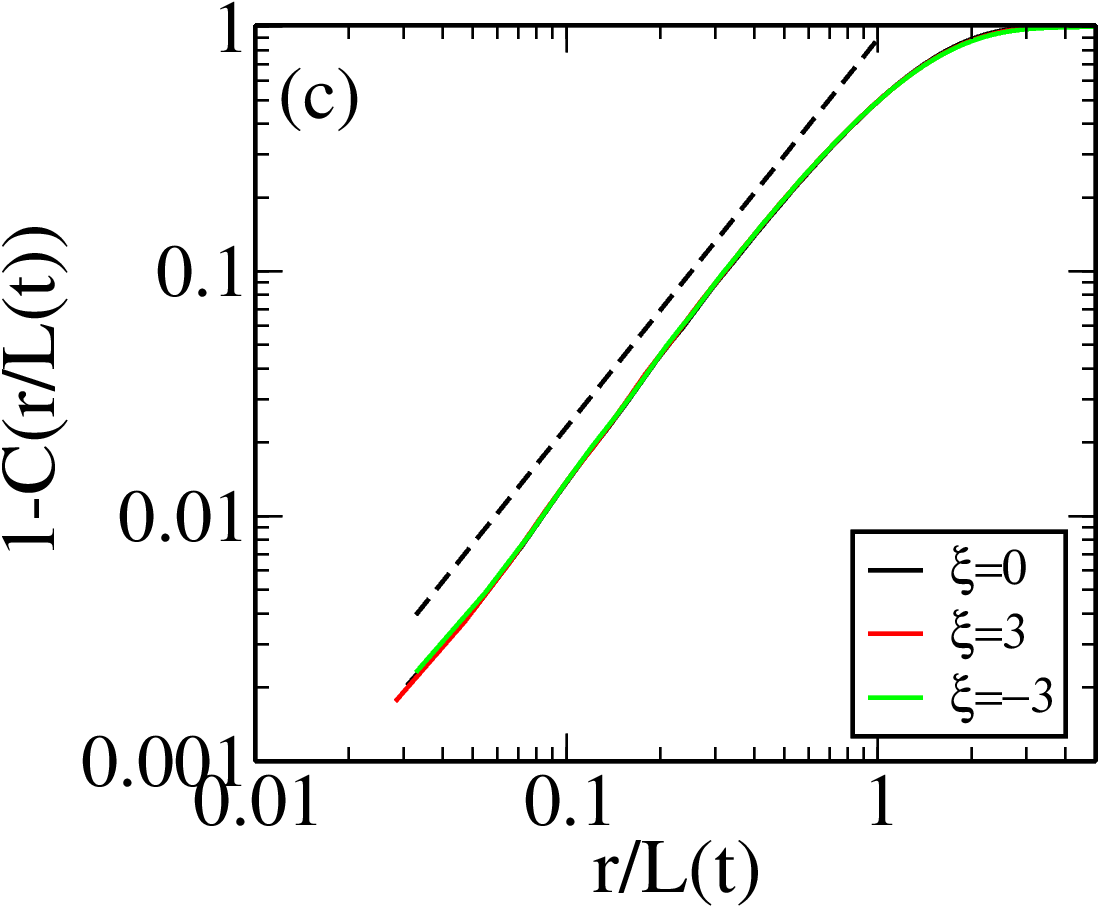}
 
	\caption{(color online) Static scaled two-point  correlation $C(r/L(t))$ {\em vs.} scaled distance $r/L(t)$ for $\xi$= $1$ , $2$, $3$ and $5$ in (a) and  $-1$, $-2$ and $-3$ in (b) for system size $L=1024$  at fixed $t=1500$. (c) Plot of $(1-C(r/L(t))$ vs. $r/L(t)$ on $\log-\log$ scale for different $\xi$. The dashed line has slope $1.6$ }
\label{fig:fig5}
\end{figure*}

{\it Growth Law}:- We characterise the domain growth by examining the growth law. i.e. the scaling of characteristic length $L(t)$ vs. time $t$ for different values and both the signs of $\xi$. In Fig. \ref{fig:fig4}(a) we show the variation of characteristic length $L(t)$'s for different $\xi$ {\em vs.} time $t$ on linear scale. The solid curve in Fig. \ref{fig:fig4}(a) is for the {\em Passive Model A}, whereas the curves on top of it are for extensile case, $\xi > 0$ and below are for the contractile case, $\xi <0$. Very clearly the characteristic length $L(t)$ decreases on decreasing $\xi$. In the Fig. \ref{fig:fig4}(b) we show the same plot on the logarithmic scale. The top and bottom dashed lines have slope $1/2$. Clearly for higher magnitude of $\xi$ the curves become closer to $L(t) \sim t^{1/2}$.
Hence hydrodynamic effect in active polar fluid does not affect the asymptotic growth law as found for the nonconserved {\it Model A} \cite{Hohenberg1977}. It only includes a correction factor. Next we give the recipe to estimate the correction factor. \\
                           
                            In Fig. \ref{fig:fig4}(c) we plot the $t/L^2(t)$ vs. $\ln(t)$ for different $\xi$ values, to  compare the results with the domain growth of  nonconserved two-component vector order parameter field  in two-dimensions, where  $L(t) \sim (t/\ln(L(t))^{1/2}$ \cite{Bray1990}. Hence $t/L^2(t)$ should vary linearly with $\ln(t)$. Which is the case for passive limit $\xi=0$ (linear variation of black solid curve) in 
                            Fig. \ref{fig:fig4}(c). We assume the deviation from the linear dependence or from the {\em Passive Model A} as a prefactor. The  approximated form of characteristic length for finite $\xi$ is 
                            $L(\xi, t)=L_{0}(\xi, t)\big(t/\ln(t)\big)^{1/2}$, where the correction factor $L_0(\xi, t)$ is obtained by following procedure. \\
                            If the growth of domain remains the same as for the {\em Passive Model A}, then the plot of $t/L^2(t)$ will be linear in $\ln(t)$. We find that for all
                            $\xi$ values for the early times $t/L^2(t)$ plot varies linearly with $\ln(t)$. It remains linear for late time for passive case $\xi=0$ and smaller activities $|(\xi)| <3$. Using the expression for the $L(t)$ we can rewrite $t/L^2(t)=\frac{\ln(t)}{L_0^2(t, \xi)}$. For larger times and larger $|\xi| > 3$, the $t/L^2(t)$ plot saturates and becomes independent of time $t$. Hence the correction $L_0^2(t, \xi)$ should vary as $\sim (\ln(t))$. And the characteristic length simply goes as $L(t) \sim t^{1/2}$, with a constant coefficient decreases with increasing activity. For larger $\xi$ data  curve starts to converge. Hence the asymptotic growth becomes pure algebraic the same as for the nonconserved order parameter with discrete symmetry \cite{Hohenberg1977}. The larger active coupling of fluid for high activity case  breaks the rotational symmetry present in continuous vector order parameter and leads the system to behave like discrete spins of Ising type \cite{yang1952}.\\

{\it Static two-point correlation function}:- We further study the domain morphology for different activities. We  calculated the equal-time correlations in orientation order parameter. The equal-time  correlation function is defined as 
before. In Fig. \ref{fig:fig5} we show the plot of equal time scaled two-point correlation function $C(r/L(t))$ {\em vs.} scaled distance $r/L(t)$ for 
different activities and fixed time $t=1500$.  
The characteristics length  $L(t)$ is defined in the same manner.
We find that the curves  show the deviation from the data collapse  when plotted as a function of scaled distance $r/L(t)$ as shown in Fig. \ref{fig:fig3}(a-b) for the extensile and contractile case respectively. We further characterise the morphology of domains by approximating small
distance limit of the scaled two-point correlation function $C(r/L(t)) \simeq (1-(r/L(t))^{\alpha})$, where $\alpha $ is defined as the cusp exponent \cite{Bray1994}.
In the Fig. \ref{fig:fig5}(c) we calculate the cusp exponent $\alpha$,
by plotting $1-C(r/L(t))$ vs. $r/L(t)$ on $\log-\log$ scale for three different cases passive $\xi=0$ and $\xi=3$ and $-3$. Although system does not show the static scaling but for all activities domain morphology remains the same and charactrised by the cusp exponent $\alpha \sim 1.6$ and it shows the deviation from the Porod's law \cite{porod}.\\

{\em Summary:}- 
Now we summarise the work. We study the ordering kinetics of active polar gel. The  active gel is defined with collection of orientable objects supplied with active stresses and momentum damping
coming from the viscosity of bulk fluid medium. The activity  is controlled by an active stress, which cannot be derived from a free energy. The system can be contractile or extensile depending upon the sign of coupling with the orientation field. We study the growth kinetics of the orientation field, when quenched from the disordered  to the ordered state. We find that for the extensile coupling the growth is enhanced and for contractile case it is suppressed with respect to passive system but the asymptotic growth law remains the same as for the nonconserved field. The activity leads a correction to the growth law of  nonconserved vector order parameter. And the  asymptotic growth approaches  pure algebraic for large magnitude of activity. Hence the system behaves equivalent  to the scalar nonconserved order parameter field \cite{Hohenberg1977}. 
We have also studied the effect of activity on the dynamic and static scaling of orientation two-point correlation function. System shows  good dynamic and no static scaling for different activities. 
Domains morphology remains unaffected due to activity and shows a deviation from Porod's law \cite{porod}. Our results can be tested on the growth kinetics of wet polar active systems and gives a new direction to understand the effect of fluid on the kinetics of orientable objects in fluids.  \\

{\em Acknowledgement:-}
S.D. acknowledges the support and the resources provided by PARAM Shivay   Facility under the National Supercomputing   Mission, Government of India at the Indian Institute of Technology, Varanasi. S.M. thanks DST-SERB India, MTR/2021/000438, andCRG/2021/006945 for financial support.

\end{document}